# Weak coupling interactions of colloidal lead sulphide nanocrystals with silicon photonic crystal nanocavities near 1.55 μm at room temperature

Ranojoy Bose[(a)], Xiaodong Yang, Rohit Chatterjee, Jie Gao, and Chee Wei Wong

*Optical Nanostructures Laboratory, Columbia University, New York, New York 10027*

Abstract:

We observe the weak coupling of lead sulphide nanocrystals to localized defect modes of 2-dimensional silicon nanocavities. Cavity resonances characterized with ensemble nanocrystals are verified with cold-cavity measurements using integrated waveguides. Polarization dependence of the cavity field modes is observed. The linewidths measured in coupling experiments are broadened in comparison to the cold-cavity characterization, partly due to large homogeneous linewidths of the nanocrystals. The calculated Purcell factor for a single exciton is 75, showing promise toward applications in single photon systems. These novel light sources operate near 1.55 μm wavelengths at room temperature, permitting integration with current fiber communications networks.

(a) rb2261@columbia.edu

The study of cavity quantum electrodynamics (CQED) in wavelength-scale optical cavities is of central interest in the field of optics and solid-state physics, and has traditionally been performed with quantum dots (QD) formed through self-assembly[1-4]. Self-assembled quantum dots (SAQD), however, are typically in III-V semiconductors and cannot benefit from the vast and advanced silicon foundry infrastructure for large-scale integration. For silicon photonic crystals, an alternative approach is to integrate colloidal QDs after device fabrication, using spin- or drop-casting techniques[5] as a post-processing step. Colloidal lead salt nanocrystal quantum dots have recently emerged as excellent candidates for charge-based as well as optical applications. The ability to synthesize these nanoparticles in various core-shells and for a variety of wavelength ranges allows for wide-ranging applications, for example, as efficient fluorescent tags in biomolecules[6]. Lead sulphide (PbS) QD, used in the experiments presented here have been studied in detail in literature[7]. Colloidal emitter-cavity interactions have recently been reported in nanocrystal-AlGaAs cavities at shorter wavelengths[8], microcapilliary resonators[9], and nanowire-1D photonic crystal structures[10]

The theory of cavity quantum electrodynamics is documented in great detail for the case of QDs in (or near) a photonic crystal cavity. In the case of weak coupling, and under perfect spectral and spatial alignment, the spontaneous emission rates of resonant excitons are modified according to the well- known Purcell factor $F_p$[11]:

$$F_p = \frac{3\lambda^3}{4\pi^2 n^3}\frac{Q}{V}$$

where $\lambda$ is the resonant wavelength of the cavity mode and $V$ the interacting cavity modal volume. The enhancement in spontaneous emission (SE) for a QD that is polarization-matched to the cavity field mode is given by the expression[8,12]:

$$E = F_p \times \frac{\gamma_c(2\gamma_e + \gamma_c)}{4(1-\frac{\omega_c}{\omega_e})^2 + (2\gamma_e + \gamma_c)^2} \times \frac{|\vec{E}_r|^2}{|\vec{E}_{max}|^2} + F_{PhC}$$

where $\omega_c$ and $\omega_e$ are the frequencies of the cavity resonance and emitter respectively, and $\gamma_e$ and $\gamma_c$ are the emitter and cavity decay rates respectively. $F_{PhC}$ is usually an inhibition induced by the photonic crystal lattice.

In these experiments, optimized *L3* nanocavities[13] in silicon are used with an additional center hole for studying both mid-cavity- plane and evanescent coupling of the PbS QD. The design parameters are as follows: *a= 420 nm, r= 0.29a (± 10%), r$_c$= 0.28a (± 10%)*, and *t= 0.6a* where *a, r, r$_c$,* and *t* represent the lattice parameter, radius of holes, radius of additional center hole, and slab thickness respectively ($r_c$= 0.26a (± 10%) is also used). All cavities are *s1*- or *s3*- detuned[13]. The silicon photonic crystal devices are fabricated on a $SiO_2$ cladding and incorporate waveguides (Fig. 1, 2) that allow for lensed optical fiber-waveguide coupling and characterization of the nanocavities using radiation collection measurements[13]. The devices are fabricated either at the Columbia facilites using e-beam lithography, or the Institute of Microelectronics Singapore using deep UV lithography. SEM images show a high quality of fabrication (Fig. 2). The silicon devices are characterized (Fig. 1) using 3D finite-difference time-domain (FDTD) simulations, using a software package with subpixel smoothing for increased accuracy[14]. A mode volume of approximately 0.07 $\mu m^3$ is calculated. An overall collection efficiency of 11% is computed for the cavity field mode using the numerical aperture (N.A.= 0.85) of the objective lens used in experiments and the simulated field profile. A collection efficiency of 8% is estimated for PbS QD in a PMMA thin-film. The designed cavity corresponds to a maximum theoretical Purcell factor of 75. SE enhancements (E) for single exciton states are estimated using the spatial distribution of the 3d electric field profile, and are modified from $F_p$ due to spatial and spectral mismatch. Enhancements are computed through a statistical distribution of QD, assuming random exciton polarization, to represent the actual measurements as well as to determine the viability of these devices in low-QD number or single photon operational regimes. Using the collection efficiencies described above, and an estimated QD density of $10^3$ per $\mu m^2$, an average overall enhancement of 1.1351 (σ: 0.1105) is calculated for weakly coupled dots for an assumed $F_{PhC}$ of 0.6 [8]. However, this prediction is limited, especially in the case of high pump-power cavity-mapping using ensemble QD, since significant sources of enhancement such as exciton-linewidth evolution[15], and QD surface proximity effects[16] are not quantitatively included.

In the experiments, ensemble PbS nanocrystals are used as a broad-band light source to decorate the resonant mode(s) of the 2D silicon photonic crystal resonator. The nanocrystals are obtained in a mixture of PMMA (5-15% by weight) and toluene (85-95% by weight) through Evidot Technologies. The nanocrystals exhibit high photoluminescence (PL) efficiency, room temperature stability, and PL peak around 1500 nm, with a full-width half-maximum of 150 nm. After diluting the commercially obtained sample 2:3 parts by volume in toluene, an overall thin-film of approximately 100 nm is achieved at a spin rate of 5000 rpm. For larger film thicknesses, radiation from the cavity-coupled nanocrystals is covered under background PL from the uncoupled nanocrystals. The 100-nm thin-film of PMMA (n=1.56) changes the band structure of the photonic crystal device[8] and shifts the cavity resonance as well as the spatial electric field profile of the cavity mode due a changed index contrast, but these changes can be effectively monitored experimentally due to the presence of waveguides on the devices, enabling cold-cavity characterization. The spectra collected from the radiation of cavity field modes with a thin-film of PMMA are shown in Fig. 3a. For some devices, excess PMMA is also selectively removed from regions away from the cavity (PMMA left only in a 9 $\mu m^2$ rectangular region around the cavity) using electron-beam lithography to suppress background PL in measurements. The operation of these latter devices is expected to be similar to the theoretical devices explored above.

The experiments are performed in two steps. In step 1, coupling measurements are performed. PbS nanocrystals located near the cavity are excited off-resonance using a pulsed Ti:S laser operating at 800 nm with a repetition rate of 80 MHz and pulse duration of 150 fs. The pump signal reaching the nanocrystals is attenuated, and the pump fluence after focusing is approximately 10 $\mu J/cm^2$. The PbS QD are found to be stable under continuous, intense illumination over a period of hours, and the experiments are repeatable over a period of several days, showing that degradation does not occur due to the laser[9]. The laser light is reflected by a high-pass filter and a 60X objective lens is used to focus the beam. The radiation from the cavity is collected with the same objective lens from a spot about 2 $\mu m$ in diameter, dispersed by a 32 cm JY Horiba Triax 320 monochromator, and detected using a liquid-nitrogen cooled Ge detector. An additional high-pass filter is used near the monochromator slit to filter out any signal from the Ti:S laser.

In step 2, waveguide characterization of the cold-cavity modes is performed in the same setup by using a tapered lens fiber butt-coupled to an on-chip waveguide. The chip is mounted vertically on a wide-range translation stage that allows for monitoring at the cavity, as well as the chip-edge for waveguide to tapered-fiber alignment, by the same objective lens (Fig. 2c). In this case, an ANDO tunable laser source operating between 1480 nm and 1580 nm at 8 dBm peak power is used, and the cavity radiation is collected from the top using the objective lens. Cavity *Q* of between 500 and 1000 is estimated by fitting Lorentzians to the experimental cold-cavity radiation spectrum for different cavity designs, after the QD have been spin-coated (Fig. 3a). In this step, the QD are not excited by the low power source, and no broadband PL is observed.

The collection path in step 1 is set up by aligning to the cavity radiation using an IR camera and a broadband laser source for fiber to waveguide excitation, as in the cold-cavity measurements. Once this path is established, the Ti:S laser is used to pump the nanocrystals for the coupling experiments. Fig. 3b shows the results of the coupling measurements. Enhancement over the background PL is observed at the cavity, compared to PL approximately 10 μm away from the cavity, where the spectrum follows the familiar Gaussian lineshape with a full-width half-maximum of around 100 nm. The spectrum in Fig. 3c shows the cavity field mode normalized to the background PL. The measurements in step 1 are confirmed with fiber-based characterization of the devices in step 2 above. However, in the coupling measurements, the individual peaks in the cold-cavity radiation spectrum (Fig. 3a) are not resolved, and a broader enhanced peak is observed, centered at a cavity mode. For different samples, and depending on whether the PMMA is selectively removed, the coupled resonances are seen to shift and are verified with the corresponding cold-cavity measurements. The broadening is partly attributed to the large expected homogeneous linewidths of coupled PbS nanocrystals. Owing to low collected signal levels from the QD, the best experimental resolution is limited to 5 nm. As an added verification that the peak is due to coupling of the QDs to the cavity, a linear polarizer is introduced in the collection path, and the collected modes show strong polarization dependence (Fig. 3d). A polarization ratio of 1.7 is inferred from the polarization extinction measurements. The observed enhancement for QD at the cavity resonance is caused due to SE enhancement as well as the higher collection efficiency of the cavity field mode compared to uncoupled QD, and matches well with theory.

In silicon-based photonic crystal cavity systems, the coupling of colloidal PbS nanocrystals to silicon photonic crystals at the near-infrared and at room temperature is experimentally demonstrated. The theoretical work presented here shows that spontaneous emission enhancements of 75 can be optimally achieved in these systems. The operation of the coupled nanocavity-nanocrystal system in silicon at around 1550 nm is especially promising because of the possibility of a single photon source that can be integrated into the present fiber infrastructure and the scalability with silicon CMOS foundries. This is an alternative to the remarkable CQED experiments performed and suggested elsewhere [17, 18].

The authors acknowledge funding support from the Dept. of Mechanical Engineering., the shared experimental facilities and cleanroom that are supported by the MRSEC and NSEC, NSF and NYSTAR programs. Some devices were fabricated at IME, from which we acknowledge the support of Dr. Dim-Lee Kwong and Dr. Mingbin Yu. The authors also thank helpful discussions with Dr. T. Yoshie, D. Englund, and Dr. D. V. Talapin, and assistance from S. Mehta.

Figure Captions

FIGURE 1. (a) FDTD visualization of silicon photonic crystal with air holes, with integrated waveguide for fiber-based characterization, showing $E_x$ field profile of cavity mode at slab center; (b) $|E_x|^2$ at slab surface.

FIGURE 2. (a) and (b) Scanning electron microscope images of fabricated silicon devices. (a) *s3* with lateral detunings of 0.176*a*, 0.025*a* and 0.176*a* for three holes adjacent to the cavity, center hole of radius 0.308*a*, (b) Angled view of typical SOI device. Scale bar represents 2 μm in (a) and 1 μm in (b). *r* = 0.319*a*. (c) Schematic of the imaging/characterization system. A Ti:S pump is used for the PL measurements to determine QD-cavity coupling (step 1: solid lines). For cold-cavity characterization (step 2: dotted lines), a tunable laser source (TLS) is coupled to the chip through lensed fibers (LF) and integrated waveguides. The samples are mounted on XYZ translation stages and the radiation collected through a 60X objective lens (OL), telescope system, and long pass filter (LPF).

FIGURE 3. (a) Cold-cavity modes characterized using tunable laser after 100 nm PMMA coating for (i) *L3*, *s3* detuning, $r_c$= 0.308*a*; $\lambda_0$=1548 nm; (ii) *s1* detuning, $r_c$ = 0.308*a*, after selective PMMA removal; $\lambda_1$=1530 nm, $\lambda_2$= 1534 nm; and (iii) *s1* detuning (.176*a*), $r_c$= 0.286*a*; $\lambda_3$=1543 nm, and $\lambda_4$=1548 nm; *r* = 0.319*a*.

(b) Coupling measurements (y-axis arbitrarily shifted): (I) background PL; (II)-(IV) coupling measurements corresponding to cold-cavity characterization in Figures 3 a-i to a-iii respectively. Measured coupled resonances at (II) $\lambda_0$=1550 nm; (III) $\lambda_1$=1535 nm; and (IV) $\lambda_3$=1545 nm; (c) Normalized cavity spectrum for the cavity in Fig. 3b-II.

(d) Polarization coupling measurements for *s1*, $r_c$= 0.308*a*. $\lambda_5$=1535 nm.

Figure 1

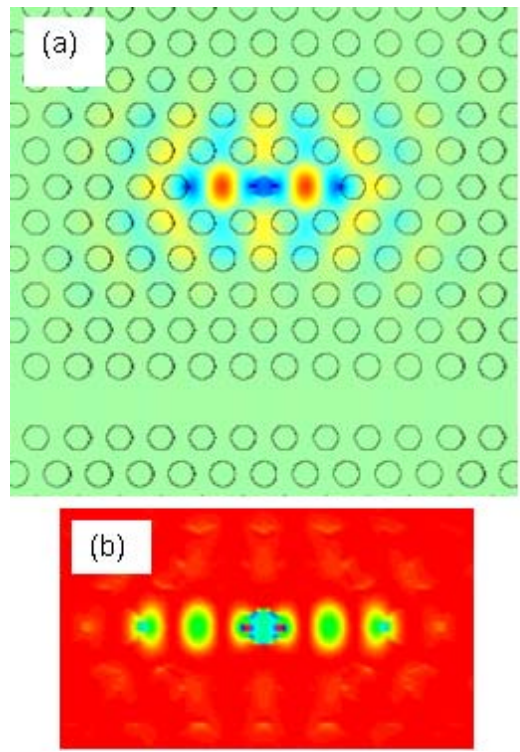

Figure 2

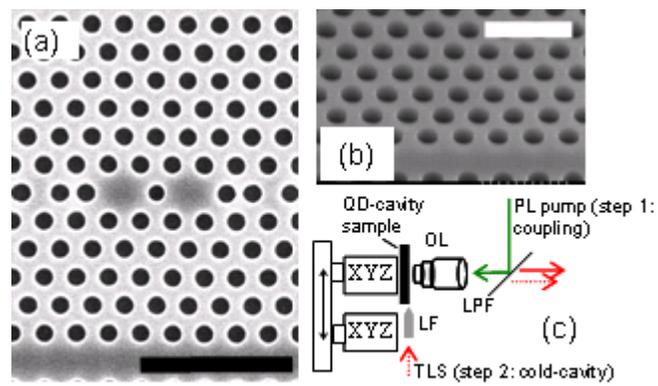

Figure 3

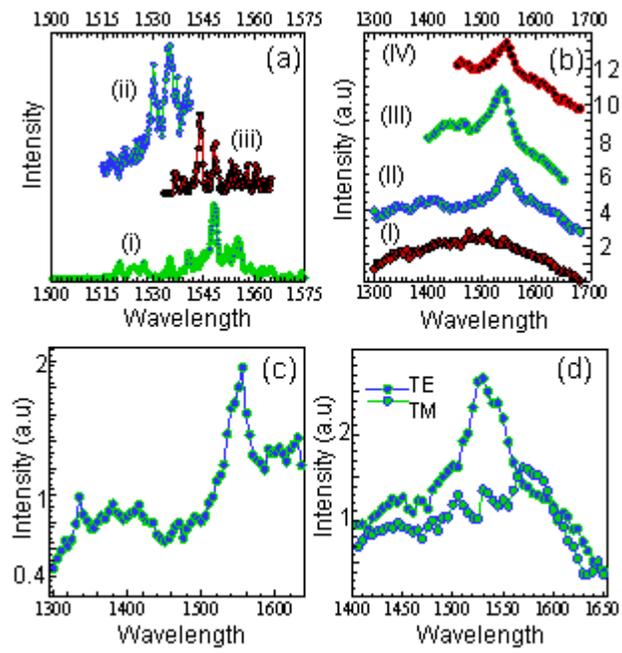